\documentclass[preprint, amsmath,amssymb,aps,prapplied,longbibliography]{revtex4-1}
\usepackage{graphicx}
\usepackage{dcolumn}
\usepackage{amsmath}
\usepackage{here}
\usepackage{bm}
\usepackage{color}

\begin{document}

\title{
Chaos and relaxation oscillations in spin-torque windmill neurons
}

\author{R. Matsumoto$^{1}$}\email{rie-matsumoto@aist.go.jp}
\author{S. Lequeux$^{2,3}$}
\author{H. Imamura$^{1}$}
\author{J. Grollier$^{2}$}

\affiliation{ 
$^{1}$National Institute of Advanced Industrial Science and Technology (AIST),
Spintronics Research Center, Tsukuba, Ibaraki 305-8568, Japan
}
\affiliation{ 
$^{2}$Unit\'{e} Mixte de Physique, CNRS, Thales, Univ. Paris-Sud, Universit\'{e} Paris-Saclay, 91767 Palaiseau, France
}
\affiliation{ 
$^{3}$Present affiliation: Univ. Grenoble Alpes, CEA, CNRS, Grenoble INP, INAC-SPINTEC, 38000 Grenoble, France
}

\date{\today}

\begin{abstract}
Spintronic neurons which emit sharp voltage spikes are required for the realization of hardware neural networks enabling fast data processing with low-power consumption. 
In many neuroscience and computer science models, neurons are abstracted as non-linear oscillators. 
Magnetic nano-oscillators called spin-torque nano-oscillators are interesting candidates for imitating neurons at nanoscale. 
These oscillators, however, emit sinusoidal waveforms without spiking while biological neurons are relaxation oscillators that emit sharp voltage spikes. 
Here we propose a simple way to imitate neuron spiking in high-magnetoresistance nanoscale spin valves 
where both magnetic layers are free and thin enough to be switched by spin torque. 
Our numerical-simulation results show that the windmill motion induced by spin torque in the proposed spintronic neurons gives rise to spikes 
whose shape and frequency, set by the charging and discharging times, can be tuned through the amplitude of injected dc current. 
We also found that these devices can exhibit chaotic oscillations. Chaotic-like neuron dynamics has been observed in the brain, 
and it is desirable in some neuromorphic computing applications whereas it should be avoided in others. 
We demonstrate that the degree of chaos can be tuned in a wide range by engineering the magnetic stack and anisotropies and by changing the dc current. 
The proposed spintronic neuron is a promising building block for hardware neuromorphic chips leveraging non-linear dynamics for computing.
\end{abstract}

%\pacs{75.78.-n, 72.25.-b}
%\keywords{spintronics, neuromorphic computing, spin torque}

\maketitle

%========================================
% Introduction
%========================================
\section{INTRODUCTION}
Neuromorphic chips need several millions of neurons to run state of the art neural networks
\cite{merolla_million_2014}.  
Keeping theses chips small therefore requires developing nanoscale artificial neurons. 
In many neuroscience and computer science models, neurons are abstracted as non-linear oscillators
\cite{hoppensteadt_oscillatory_1999, aonishi_statistical_1999, jaeger_harnessing_2004, maass_real-time_2002}.   
%[F. C. Hoppensteadt, and E. M. Izhikevich, PRL (1999)] [T. Aonishi,  PRL (1998)]
%[H. Jaeger, Science (2004)] [W. Maass, Neural Comput. (2002)]. 
Memristive oscillators (also called neuristors) 
\cite{pickett_scalable_2013},
%[M. D. Pickett, Nat. Mater. (2013)], 
Josephson junctions 
\cite{segall_synchronization_2017},
%[K. Segall, Phys. Rev. E (2017)], 
nanoelectromechanical systems
\cite{feng_self-sustaining_2008}, 
%[X. L. Feng, Nat. Nanotech. (2008)] 
and magnetic nano-oscillators called spin-torque nano-oscillators 
\cite{slonczewski_current-driven_1996, berger_emission_1996, kiselev_microwave_2003} 
%[J.C. Slonczewski, JMMM (1996)] [L. Berger, PRB (1996)] [S. I. Kiselev, Nature (2003)] 
are interesting candidates for imitating neurons at the nanoscale. 
In particular, it has been shown experimentally that spin-torque nano-oscillators can implement hardware neural networks 
and perform cognitive tasks with high accuracy due to their large signal to noise ratio, 
their high non-linearity and enhanced ability to synchronize 
\cite{torrejon_neuromorphic_2017}.
%[J. Torrejon, Nature (2017)]. 

However, the microwave voltage signals delivered by these spin valves driven by spin torque are typically sinusoidal. 
In contrast, biological neurons are relaxation oscillators, 
based on two time scales: a long charging period followed 
by a short discharge period 
\cite{pikovsky_synchronization_2003, buzsaki_rhythms_2011}.
%[A. Pikovsky, Synchronization, Cambridge University Press (2001)] [G. Buzsaki, Oxford Univ. Press, (2011)]. 
Their output consists of sharp voltage spikes of fixed amplitude with a frequency that depends on the amplitude of the inputs. 
Therefore, it is interesting to exploit the multifunctionality and tunability of spin-torque to imitate the sharp neuron spikes. 

%==============================
% Fig. 1
%==============================
\begin{figure}[H]
  \includegraphics [width=0.95\columnwidth] {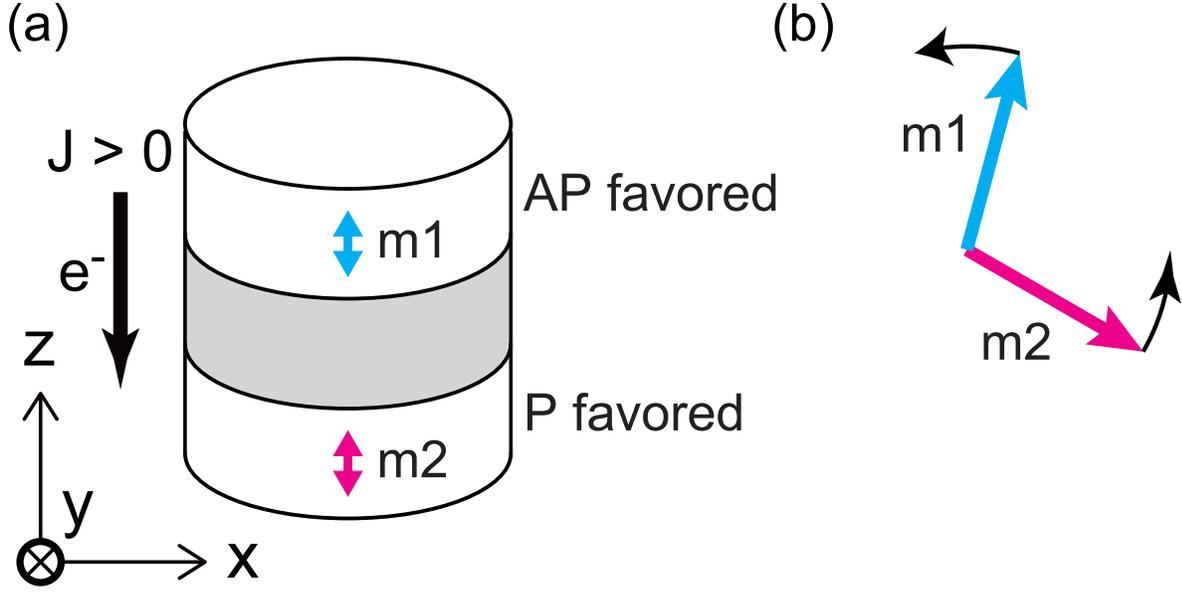}%
  \caption{
  \label{fig:fig1} 
(a) Schematic of the windmill neuron: a spin valve with two free layers. 
The double-headed arrow in cyan (magenta) represents the magnetization unit vector, m1 (m2), 
in the equilibrium states. The axis $z$ is parallel to the out-of-plane (OOP) direction. 
When the current density ($J$) is positive, electrons (e$^{-}$) flow from m1 to m2. 
(b) Schematic of the windmill spin-torques configuration of the two magnetizations m1 and m2. 
In the case of $J>0$, m1 switches away from m2, and m2 follows m1 as indicated by the arrows.
In other words, at $J>0$, m1 favors antiparallel (AP) configuration, 
and m2 favors parallel (P) configuration, as indicated in (a). 
  }
\end{figure}

%========================================
% In this paper
%========================================
Here we propose a simple way to imitate neuron spiking in high-magnetoresistance nanoscale spin valves 
where both magnetic layers are free and thin enough to be switched through spin torque 
\cite{gupta_switching_2016, choi_current-induced_2016, thomas_spin_2017}.
%[G. Gupta, arXiv (2016), choi_current-induced_2016,  L. Thomas, INTERMAG (2017)]. 
We study these devices through macrospin and micromagnetic simulations 
\cite{Spin-PM}.
%[SpinPM]. 
We show that the windmill motion induced by spin torque 
\cite{slonczewski_theory_2007}
%[J. C. Slonczewski, J. Z. Sun, JMMM (2007)] 
in these structures gives rise to spikes whose shape and frequency, set by the charging and discharging times, 
can be tuned through the amplitude of injected dc current as well as the materials and thicknesses of the ferromagnetic layers. 
We observed that these devices with many coupled degrees of freedom can exhibit chaotic oscillations. 
Chaotic-like neuron dynamics has been observed in the brain 
\cite{softky_highly_1993},
%[W. R. Softky, J Neurosci. (1993)], 
and is desirable in some neuromorphic computing applications 
\cite{kumar_chaotic_2017}
%[S. Kumar, Nature (2017)] 
whereas it should be avoided in others 
\cite{appeltant_information_2011}.
%[L. Appeltant, Nat. Commun. (2011)]. 
We point out that the dipolar coupling between magnetic layers is the main source of chaos in spin-torque windmill neurons. 
We demonstrate that the degree of chaos can be tuned in a wide range by engineering the magnetic stack and anisotropies. 
The proposed spiking windmill spin-torque neuron with controllable chaos is a promising building block 
for hardware neuromorphic chips leveraging non-linear dynamics for computing.

%========================================
% Principle
%========================================
\section{Windmill relaxation oscillations: principle}
The structure of the proposed windmill neuron, illustrated in Fig. 1(a), is a spin valve, 
consisting of a nonmagnetic spacer layer sandwiched between two ferromagnetic layers. 
The spacer layer can be either a metallic layer in giant magnetoresistance devices 
\cite{baibich_giant_1988, binasch_enhanced_1989},
%[M. N. Baibich, PRL (1988)] [G. Binasch, PRB 39 4828 (1989)], 
or a thin insulating tunnel barrier layer in magnetic tunnel junctions 
\cite{miyazaki_giant_1995, moodera_large_1995, yuasa_giant_2004, parkin_giant_2004, djayaprawira_230_2005}.
%[T. Miyazaki, JMMM (1995)] [J. S. Moodera, PRL (1995)] [S. Yuasa, Nat. Mater. (2004)] [S. S. P. Parkin, Nat. Mater. (2004)] [D. D. Djayaprawira, APL (2005)]. 
The two magnetizations, m1 and m2, have preferential directions due to magnetic anisotropy. 
However, contrary to typical spin-valve stacks, both layers are free to switch: none of them is pinned. 
In the absence of spin torque, the magnetization directions are either parallel (P) or antiparallel (AP). 
They can point in-plane (IP) 
\cite{myers_current-induced_1999}
or out-of-plane (OOP)
\cite{mangin_current-induced_2006}, 
depending on the dominant source of anisotropy. 
When a dc current is injected in the spin valve, perpendicularly to the layer planes, 
the torques on the two magnetizations tend to induce rotations in the same direction, as illustrated in Fig. 1(b). 
The direction of rotation is set by the sign of the applied dc current.

\section{MODEL}
%========================================
% Model
%========================================
It has been predicted, as well as experimentally observed 
that this torque configuration can generate a windmill-like motion of the two magnetizations 
\cite{gupta_switching_2016, choi_current-induced_2016, thomas_spin_2017}.
%[G. Gupta, arXiv (2016), R. Choi, IEEE Trans. Magn. (2016),  L. Thomas, INTERMAG (2017)] [J. C. Slonczewski, J. Z. Sun, JMMM (2007)]. 
The equations of motion of the magnetizations are given by the Landau-Lifschitz-Gilbert-Slonczewski (LLGS) equation 
\cite{slonczewski_current-driven_1996, berger_emission_1996, stiles_spin-transfer_2006}$\colon$ 
%[J.C. Slonczewski, JMMM (1996)] [L. Berger, PRB (1996)] [Stiles_spin-transfer_2006]

\begin{align}
  \label{eq:LLG1}
\frac{\partial \textbf{\textit{m}}_{1} }{\partial t}
    = -\gamma \textbf{\textit{m}}_{1} \times &
    \textbf{\textit{H}}_{\rm eff1}
    +\alpha
    \textbf{\textit{m}}_{1} \times
    \frac{\partial \textbf{\textit{m}}_{1} }{\partial t} \nonumber\\
    &-\gamma \tau_{\rm st1} \textbf{\textit{m}}_{1} \times (\textbf{\textit{m}}_{1} \times \textbf{\textit{m}}_{2}), \\
   \label{eq:LLG2}
\frac{\partial \textbf{\textit{m}}_{2} }{\partial t}
    = -\gamma \textbf{\textit{m}}_{2} \times  &
    \textbf{\textit{H}}_{\rm eff2}
    +\alpha
    \textbf{\textit{m}}_{2} \times
    \frac{\partial \textbf{\textit{m}}_{2} }{\partial t}  \nonumber\\
    &+\gamma \tau_{\rm st2} \textbf{\textit{m}}_{2} \times (\textbf{\textit{m}}_{2} \times \textbf{\textit{m}}_{1}). 
\end{align}
Here, $t$ and $\gamma$ are the time and the electron gyromagnetic ratio. The second term on the right-hand side of 
Eqs. (\ref{eq:LLG1}) and (\ref{eq:LLG2}) is the damping-torque term 
where $\alpha$ is the Gilbert damping constant. 
In this article, $\alpha=0.01$ is assumed. Hereafter, $i$ in the subscript represents the quantities of m$_{i}$ layer with $i = 1$ or 2. 
$\tau_{st1}$, and $\tau_{st2}$ represent the coefficient of the Slonczewski torque$\colon$
\begin{align}
  \label{eq:tau}
\tau_{sti} = \frac{\hbar}{2} \frac{1}{\mu_{0} M_{\rm s}} \frac{1}{d_{i}} \frac{J}{|e|} P.
\end{align}
Here, $\hbar$ is the Dirac constant, $\mu_{0}$ is the vacuum permeability, 
$M_{s}$ is the saturation magnetization, $d_{i}$ is the thickness of layer i, 
$e$ is the electron charge, $J$ is the current density and $P$ is the spin polarization. 
In the rest of the article, we take $P=0.6$.

$\textbf{\textit{H}}_{{\rm eff}i}$ is the effective field expressed as
\begin{align}
  \label{eq:Heff}
\textbf{\textit{H}}_{\rm eff} = \textbf{\textit{H}}_{\rm anis} + \textbf{\textit{H}}_{\rm dip}.
\end{align}
Hereafter the layer index, $i$, is abbreviated.
$\textbf{\textit{H}}_{\rm anis}$ represents the anisotropy field expressed as$\colon$
\begin{align}
  \label{eq:Hanis}
\textbf{\textit{H}}_{\rm anis}=\frac{2K}{\mu_{0}M_{\rm s}} m_{z}. 
\end{align}
Here $K$ represents the anisotropy constant. 
In the spin valve shown in Fig. 1(a), $K = 115$ kJ/m$^{3}$ is assumed in m1, and $K = 70$ kJ/m$^{3}$ is assumed in m2. 
$\textbf{\textit{H}}_{\rm dip}$ represents the dipolar field expressed as
\begin{align}
  \label{eq:Hdemag}
\textbf{\textit{H}}_{\rm dip}= - M_{\rm s} \left( N_{x} m_{x},  N_{y} m_{y},   N_{z} m_{z} \right). 
\end{align}
Here $N_{x}$, $N_{y}$ and $N_{z}$ are the demagnetization coefficients 
\cite{beleggia_demagnetization_2005}.
%[M. Beleggia, J. Phys. D (2005)].

%========================================
% RESULTS
%========================================
\section{RESULTS}
To highlight the principle of windmill neurons, 
we first neglect the dipolar-field interactions between the two magnetic layers in Fig. 1(a) 
and consider that they behave as macrospins with uniform magnetizations. 
%==============================
% Fig. 2
%==============================
\begin{figure}[H]
%\begin{figure}[P]
\includegraphics [width=0.8\columnwidth] {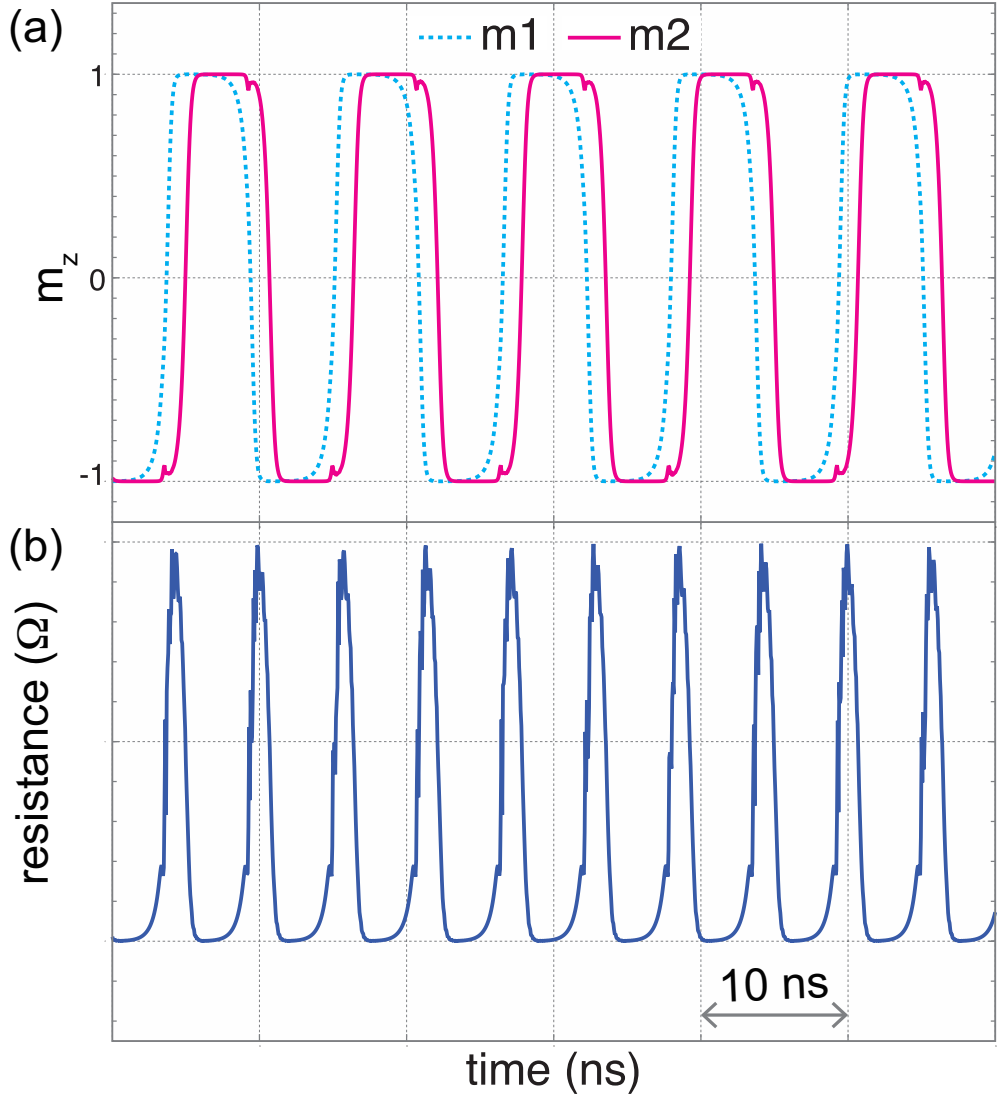}
  \caption{
  \label{fig:eContours} 
  (a) 
  Times traces of magnetization switching (macrospin simulations for a spin valve with out-of-plane (OOP)  
  magnetized layers without dipolar-field interaction between the two ferromagnetic layers, SV$_{\rm OOP1}$) 
  (b) 
  Corresponding resistance time trace. In (a) and (b), $J$ is positive and the normalized current density is $J/J_{th}=1.5$.  
}
\end{figure}
Fig. 2(a) shows macrospin simulations of magnetic switching in these conditions, 
for out-of-plane  magnetized layers that differ only through their anisotropy constants$\colon$ 
$K_{\rm m1}$ = 115 kJ/m$^{3}$ and $K_{\rm m2}$ = 70 kJ/m$^{3}$ 
(The other magnetic parameters are indicated as SV$_{\rm OOP1}$ in TABLE I). 
The windmill motion induces sustained switching of the magnetizations one after the other at $J \geq J_{th} = 1.0$ MA/cm$^{2}$ 
where $J_{th}$ is the threshold current density for sustained windmill switching. 
The repeated magnetic switches give rise to changes in the device resistance ($R$) 
through magnetoresistance (MR) effects$\colon$ 
$R =  (R_{AP}+R_{P})/2 - [(R_{AP} - R_{P})/2]\cos \hat{\theta}_{12}$ where $R_{P}$ ($R_{AP}$) is the resistance in the parallel (antiparallel) configuration 
and $\hat{\theta}_{12}$ is the angle between m1 and m2. 
Since the injected current is dc, the resulting voltage variations across the spin valve, i.e., $V = R \times I_{dc}$, 
are proportional to the resistance variations. 
In this article we always consider a MR ratio $(R_{AP} - R_{P})/R_{P}$ of 100\% and $R_{P} = 100$ $\Omega$
assuming that spin valves are magnetic tunnel junctions. 
The resistance variations corresponding to the magnetic switches in Fig. 2(a) are plotted in Fig. 2(b). 
A spiking behavior similar to neuron responses is observed. 
The time scales of these relaxation oscillations are set by the switching times of the two layers.  
Here the long charging period corresponds to the switching of m1 and the short discharge period to the switching of m2.

The asymmetry of the switching times comes from the different anisotropy constants of the layers used in the simulations 
($K_{\rm m1}$ = 115 kJ/m$^{3}$ and $K_{\rm m2}$ = 70 kJ/m$^{3}$). 
Indeed, the magnetization switching time $T_{SW}$ under spin torque is proportional to $1/(J-J_{th}^{(0)})$
\cite{sun_spin-current_2000}
%[J. Z. Sun, PRB (2000)], 
where $J_{th}^{(0)}$ is the individual threshold current density for switching 
\cite{choi_current-induced_2016}.
%[R. Choi, IEEE Trans. Magn. (2016)]. 
Layers with higher anisotropy $K$ are more difficult to be switched, and have a larger threshold current density $J_{th}^{(0)}$. 
In our case, we find through simulations that $J_{th1}^{(0)}$ and $J_{th2}^{(0)}$ are respectively equal 
to 0.95 MA/cm$^{2}$ and 0.49 MA/cm$^{2}$ 
where m2 (m1) is fixed at the equilibrium state during the evaluation of $J_{th1}^{(0)}$  ($J_{th2}^{(0)}$). 
The switching times during the windmill motion for the two magnetic layers as a function of current density are plotted in Fig. 3(a) (solid curves), 
together with the corresponding fits in $T_{SWi} = c_{i}/(J - J_{thi}')$ (dotted curve and dotted-dashed curve). 
%==============================
% Fig. 3
%==============================
\begin{figure*}
\includegraphics{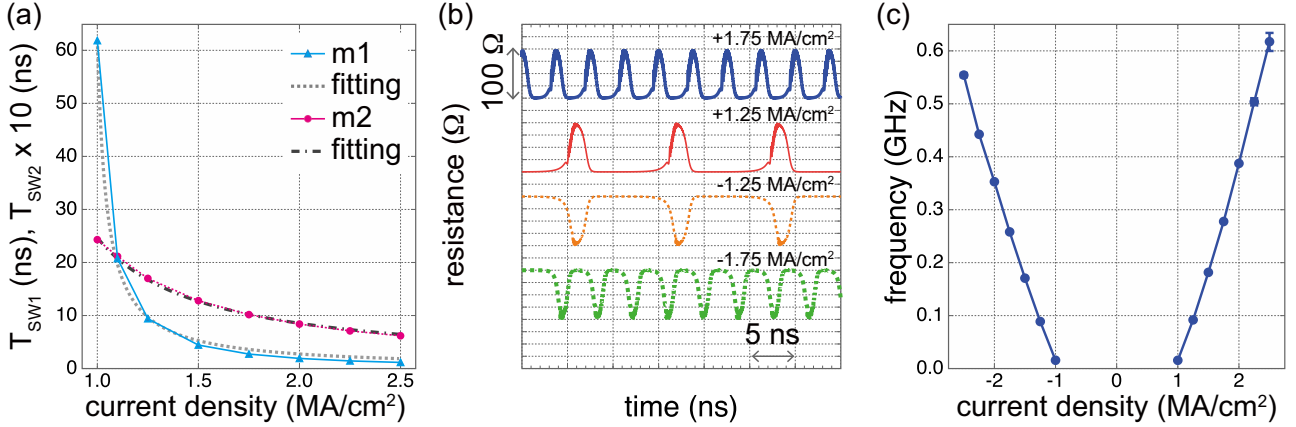}% Here is how to import EPS art
\caption{\label{fig:fig3}
(a) Average switching times ($T_{SW1}$ and $T_{SW2}$) for m1 (solid triangles on solid curve) and m2 (solid circles on solid curves) as a function of current density. 
Fits by $ci/(J-J_{thi}')$ (dotted curve and dotted-dashed curve). 
(b) Resistance times traces at $J = +1.75$ MA/cm$^{2}$ (thick solid curve), +1.25 MA/cm$^{2}$ (thin solid curve), -1.25 MA/cm$^{2}$ (thin dotted curve), -1.75 MA/cm$^{2}$ (thick dotted curve). 
(c) Frequency as a function of current density for negative and positive current densities.
}
\end{figure*}
Here, $c_{i}$ and $J_{thi}'$ are fitting parameters. The agreement between the analytical prediction of 
Ref. \cite{sun_spin-current_2000}
%[J. Z. Sun, PRB (2000)] 
and our simulations is excellent. 
The fitting yields $J_{th1}'=0.954 \pm 0.002$ MA/cm$^{2}$, $c_{1}=4.088 \pm 0.215$ ns$\cdot$MA/cm$^{2}$, 
$J_{th2}'=0.471 \pm 0.015$ MA/cm$^{2}$, and $c_{2}=1.305 \pm 0.031$  ns$\cdot$MA/cm$^{2}$. 
The threshold currents extracted from the switching times 
$J_{th1}'$ ($J_{th2}'$) 
agree well with the previously determined threshold currents $J_{th1}^{(0)}$ ($J_{th2}^{(0)}$). 
These results show that the response of the windmill neuron can be tuned by dc current. 
Traces at different dc current densities are shown in Fig. 3(b), and the evolution of the frequency as a function of current is plotted in Fig. 3(c). 
As determined experimentally and numerically in previous studies 
\cite{gupta_switching_2016, choi_current-induced_2016, thomas_spin_2017},
%[G. Gupta, arXiv (2016), R. Choi, IEEE Trans. Magn. (2016),  L. Thomas, INTERMAG (2017)], 
the frequency increases with an increase of $|J|$.  
Note that the shape of spikes can also be tuned by controlling the switching time ratio through materials engineering of the two layers ($M_{s}$, $P$ etc.).

%========================================
% Occurrence of chaos
%========================================
\section{Occurrence of chaos}
Fig. 4 compares resistance versus time traces simulated through macrospin equations of motion for in-plane (Fig. 4(a)) 
and out-of-plane magnetized spin valves (Fig. 4(b)) (the structure of the in-plane magnetized spin valve, SV$_{\rm IP1}$, 
and its parameters are shown in Fig. 6(c) and TABLEs I and II). 
As can be seen, the trace in the out-of-plane case is highly regular 
whereas apparent 
fluctuation 
affects the periodicity of switching in the in-plane case, 
even if temperature induced fluctuations are not included in the simulations.

This chaotic switching of in-plane spin valves 
\cite{montoya_magnetization_2018}
%[E. A. Montoya, arXiv:1806.03383] 
under windmill motion can be interpreted in the following way. 
For windmill motion, the switching of one layer toggles the switching of the other. 
Indeed, magnetization m1 wants to achieve the AP configuration 
whereas m2 wants to maintain a P configuration (and inversely for a reversed sign of the current density), 
therefore the P and AP configurations become consecutively unstable. 
But the switching trajectories are very different for in-plane and out-of-plane magnetized samples. 
As shown in Fig. 5(a), for in-plane magnetized samples, 
the strong anisotropy distorts the trajectories in a clamshell shape. 
Let us consider the situation 
where one of the magnetizations, m2, is close to equilibrium and the other one m1, is switching towards m2. 
The switching of m1 from one hemisphere to the other is strongly determined 
by the exact magnetization dynamics in the narrow window highlighted in Fig. 5(a). 
In this window, the angle between magnetizations $\hat{\theta}_{12}$ 
that gives the torque strength is also strongly varying. 
Therefore, small variations in the position of m2 will strongly influence the switching of m1. 
This high coupling between degrees of freedom induces a high sensitivity of magnetization reversal to initial conditions 
and can favor the appearance of chaos. 
The situation is different for out-of-plane magnetized samples, 
where precessions remain mostly circular during the whole switching of m1 (Fig. 5(b)) 
and are therefore much less sensitive to fluctuations of m2.

%==============================
% Fig. 4
%==============================
\begin{figure}[H]
%\begin{figure}[P]
\includegraphics [width=0.95\columnwidth] {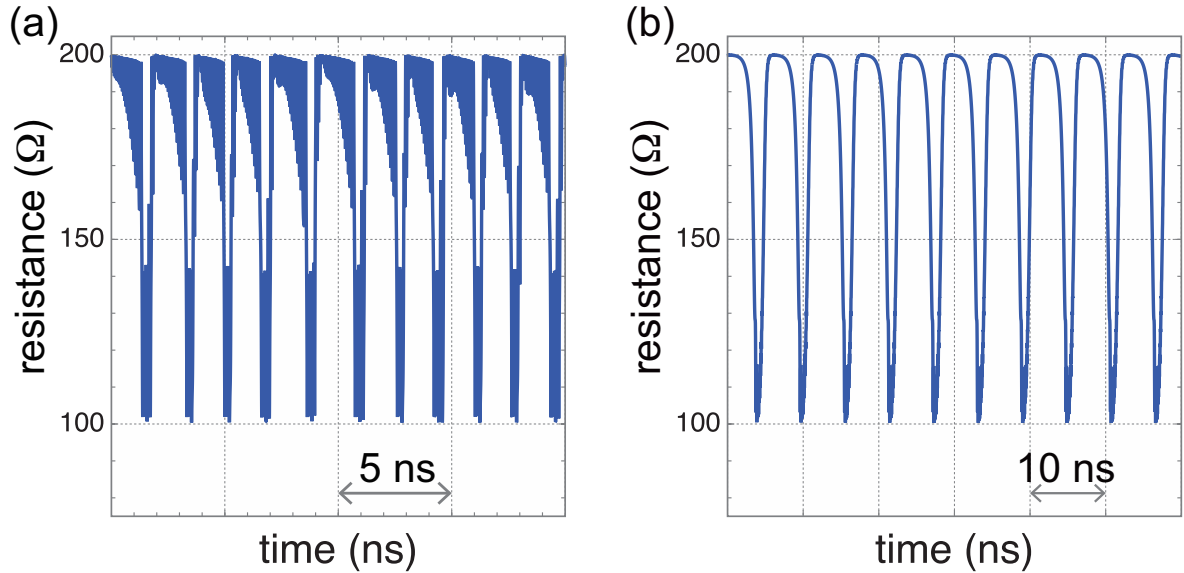}
\caption{\label{fig:fig4} 
Resistance time traces for (a) in-plane and (b) out-of-plane magnetized layers at negative $J$ with $J/J_{th}=1.5$.
}  
\end{figure}

%==============================
% Fig. 5
%==============================
\begin{figure}[H]
%\begin{figure}[P]
\includegraphics [width=0.95\columnwidth] {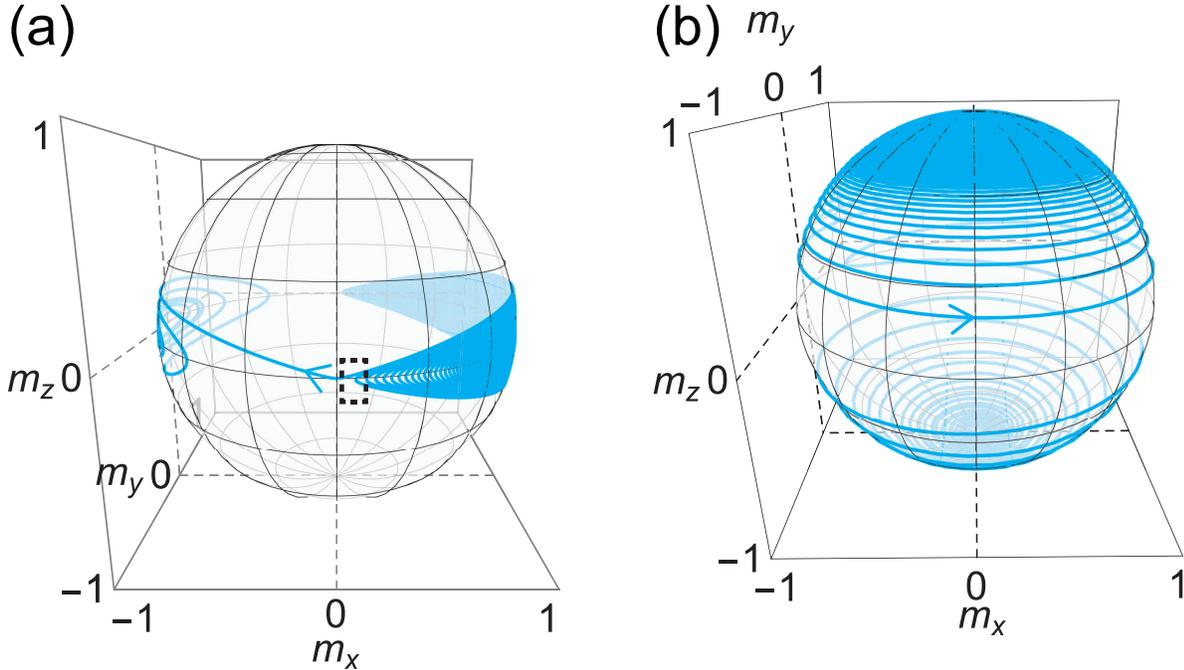}
\caption{\label{fig:fig5} 
Sketch of magnetization orbits of m1 for (a) in-plane and (b) out-of-plane magnetized layers at negative $J_{th}$. 
}  
\end{figure}

Until now we have not included the dipolar-field interaction between the magnetic layers in the simulations. 
The dipolar-field interaction 
is expected to enhance strongly the chaoticity of the system 
because it increases coupled degrees of freedom. 
Indeed, 
if the dipolar-field interaction exists,
the switching of m2 will strongly depend on the direction of m1 (and reciprocally), 
yielding an increased sensitivity of the repeated magnetization switching events on initial conditions.

%========================================
% Tuning chaos by structure
%========================================
\section{Tuning chaos by structure}
The strength of the dipolar-field interaction between layers (dipolar coupling) can be controlled 
by tuning the anisotropy and by tuning the stack. 
In this section we compare the windmill dynamics in the different structures sketched in Fig. 6.

%==============================
% Fig. 6
%==============================
\begin{figure*}
\includegraphics{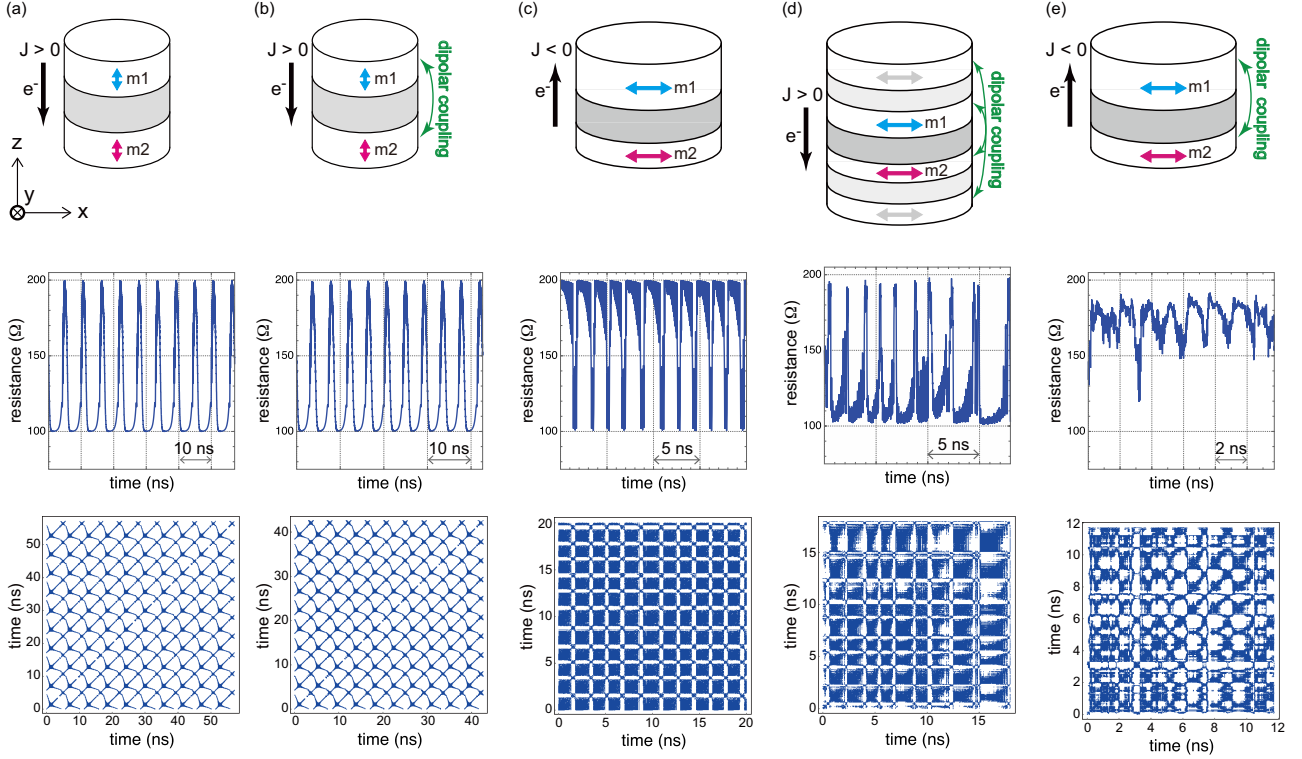}% Here is how to import EPS art
\caption{\label{fig:fig6}
Sketch of the different structures$\colon$
(a) macrospin out-of-plane (OOP) spin valve without dipolar coupling (SV$_{\rm OOP1}$), 
(b) micromagnetic OOP spin valve with dipolar coupling (SV$_{\rm OOP2}$), 
(c) macrospin in-plane (IP) spin valve without dipolar coupling (SV$_{\rm IP1}$), 
(d) micromagnetic IP spin valve with antiferromagnetically-coupled layers (m$_{\rm A1}$ and m$_{\rm A2}$) and with dipolar coupling (SV$_{\rm IP2}$), 
(e) micromagnetic IP spin valve with dipolar coupling (SV$_{\rm IP3}$). 
The axis $x$ ($z$) is parallel to the major axis of the ellipse (out-of-plane direction). 
Typical traces of the resistance versus time and the corresponding recurrence plots are shown below each case for $J/J_{th} = 1.5$. 
}
\end{figure*}

Fig. 6(a) shows an out-of-plane bilayer macrospin spin valve simulated without dipolar-field interaction (SV$_{\rm OOP1}$). 
The dipolar-field interaction is included in the micromagnetic simulations of the out-of-plane bilayer (SV$_{\rm OOP2}$) of Fig. 6(b) 
(micromagnetic simulations are described in the Appendix A). 
However, the dipolar-field interaction in the out-of-plane configuration is expected to be small because of the small $M_{s}$.  
Fig. 6(c) shows an in-plane bilayer macropin spin valve without the dipolar-field interaction (SV$_{\rm IP1}$). 
Fig. 6(d) shows a structure with a more complicated stack, where the free layers are each composed of two antiferromagnetically coupled layers (SV$_{\rm IP2}$). 
The dipolar field between the two free layers is expected to be strongly minimized in this configuration thanks to flux closure. 
Finally Fig. 6(e) shows an in-plane bilayer spin valve (SV$_{\rm IP3}$) including the dipolar-field interaction 
which is expected to be strong in this configuration. 
Because of the dipolar-field interaction, SV$_{\rm IP3}$ favors AP magnetization configuration. 
As a result, the switching from AP to P configuration is often interrupted, 
and the resistance oscillates in a higher range around 150 - 200 $\Omega$.

\begin{table*}
\caption{\label{tab:table1}
Structures under study (SV$_{\rm OOP1}$, SV$_{\rm OOP2}$, SV$_{\rm IP1}$, SV$_{\rm IP2}$, and SV$_{\rm IP3}$), simulation method, and threshold current density ($J_{th}$). 
OOP1, OOP2, IP1, IP2 are the labels of magnetic layers whose parameters are summarized in TABLE II. 
In all structures, the spacer layer between m1 and m2 has the thickness of 1 nm and its $J_{\rm RKKY}=0$. 
In (d) SV$_{\rm IP2}$, the spacer layer between m1 and m$_{\rm A1}$ (m2 and m$_{\rm A2}$) has the thickness of 0.7 nm and its $J_{\rm RKKY}=-0.1$ mJ/m$^{2}$.
}
\begin{ruledtabular}
\begin{tabular}{cccccc}
 Structure & (a) SV$_{\rm OOP1}$  &  (b) SV$_{\rm OOP2}$  &  (c) SV$_{\rm IP1}$  &  (d) SV$_{\rm IP2}$ &  (e) SV$_{\rm IP3}$  \\ \hline
 m$_{\rm A1}$  & ---  &  ---  &  ---   &  IP1  &  ---  \\
 m1    & OOP1  &  OOP1  &  IP1    &  IP1  &  IP1  \\
 m2    & OOP2  &  OOP2  &  IP2    &  IP2  &  IP2  \\
 m$_{\rm A2}$  & ---  &  ---  &  ---   &  IP2  &  ---  \\
 Simulation method & Macrospin\footnote{Macrospin-model simulations were conducted without dipolar coupling.} &  
 Micromagnetics\footnote{Micromagnetic simulations were conducted with dipolar coupling.} &  Macrospin$^{\text{a}}$ 
 &  Micromagnetics$^{\text{b}}$  &  Micromagnetics$^{\text{b}}$   \\ 
%  \begin{tabular}{c} Simulation \\ method \end{tabular}  & \begin{tabular}{c} Macrospin  without \\ dipolar coupling \end{tabular} &  \begin{tabular}{c} Micromagnetics with \\ dipolar coupling \end{tabular}  &  \begin{tabular}{c} Macrospin  without  \\ dipolar coupling \end{tabular}  
% &  \begin{tabular}{c} Micromagnetics with \\ dipolar coupling \end{tabular}   &  \begin{tabular}{c} Micromagnetics with \\ dipolar coupling \end{tabular}   \\ 
  \begin{tabular}{c} $J_{th} (>0)$ (MA/cm$^{2}$)  \\  $J_{th} (<0)$ (MA/cm$^{2}$)   \end{tabular}  &   \begin{tabular}{c} 1.0 \\  -1.0  \end{tabular}   &   \begin{tabular}{c} 1.1 \\  -1.1  \end{tabular}   
  &  \begin{tabular}{c} ---\footnote{
 In (c) SV$_{\rm IP1}$ and (e) SV$_{\rm IP3}$, positive current induces continuous spin-torque oscillations of m1 and m2 which does not result in  spiking time trace of resistance.
                                                             }\\  -6.0  \end{tabular}  
  &  \begin{tabular}{c} 21 \\  -23  \end{tabular}   
  & \begin{tabular}{c} ---$^{\text{c}}$ \\  -34  \end{tabular}  \\ 
\end{tabular}
\end{ruledtabular}
\end{table*}

\begin{table*}
\caption{\label{tab:table2}
Parameters of magnetic layers: OOP1, OOP2, IP1, and IP2. $S$ is the area of the base.
}
\begin{ruledtabular}
\begin{tabular}{ccccc}
 Magnetic layer & OOP1 & OOP2 & IP1 & IP2 \\ \hline
 $S$ (nm$^{2}$)  &  $16\times16\times\pi$  &  $16\times16\times\pi$  &  $30\times10\times\pi$  &  $30\times10\times\pi$  \\
 $d$ (nm) & 1 & 1  &  1  &  0.5 \\
 $M_{s}$ (kA/m)  &  200  &  200 &  1300 & 1300 \\
 $K$ (kA/m$^{3}$) &   115   &  70   &   0  &  0  \\
\end{tabular}
\end{ruledtabular}
\end{table*}

Typical time traces are shown below each structure. 
As can be seen, the degree of chaos seems to increase 
when the anisotropy changes from out-of-plane (Fig. 6 (a)-(b)) to in-plane (Fig. 6(c)). 
It also increases in the in-plane configuration 
when the strength of dipolar-interaction between layers increases (Fig. 6 (c)-(d)-(e)).

In order to evaluate more thoroughly the degree of chaos in structures shown in Fig. 6, 
we have used three methods: quality factor (Q factor), Recurrence Quantification Analysis (RQA) 
\cite{eckmann_recurrence_1987, marwan_recurrence_2007, webber_dynamical_1994, marwan_nonlinear_2002},
%[J. P. Eckmann, Europhys. Lett. (1987)] [N. Marwan, Phys. Rep. (2007)] 
%[C. L. Webber Jr.,and J. P.Zbilut, J Appl Physiol. (1994)] [N. Marwan and J. Kurths,  Phys. Lett. A (2002)] 
%[N. Marwan, Phys. Rep. (2007)], 
and Lyapunov exponent 
\cite{wolf_determining_1985}.
%[A. Wolf, Physica 16D (1985)]. 
Low Q factor and low $DET$, $L$, $L_{\rm max}$ and $ENTR$ in RQA indicate high degree of chaos,
and high Lyapunov exponent indicates high degree of chaos. 
The evaluated values at $J/J_{th} = 1.5$ are summarized in TABLE III.

\begin{table*}
\caption{\label{tab:table3}
Evaluated degree of chaos: quality factor (Q factor), [$DET$, $L$, $L_{\rm max}$, $ENTR$] of Recurrence Quantification Analysis (RQA), 
and Lyapunov exponent of SV$_{\rm OOP1}$, SV$_{\rm OOP2}$, SV$_{\rm IP1}$, SV$_{\rm IP2}$, and SV$_{\rm IP3}$ at $J/J_{th} = 1.5$.
}
\begin{ruledtabular}
\begin{tabular}{cccccc}
 Method & (a) SV$_{\rm OOP1}$  &  (b) SV$_{\rm OOP2}$  &  (c) SV$_{\rm IP1}$  &  (d) SV$_{\rm IP2}$ &  (e) SV$_{\rm IP3}$  \\ \hline
 Q factor  & $>10^{4}$  &  440  &  11   &  2.5  &  3.7  \\
 $DET(\%)$   &  0.88  &  0.66  &  0.10    &  0.066  &  0.11  \\
 $L$   &  5.4  &  4.5  &  2.4    &  2.2  &  2.3  \\
 $L_{\rm max}$    & 3100  &  650  &  40    &  8  &  9  \\
 $ENTR$  & 1.9  &  1.7  &  0.81   &  0.44  &  0.69  \\
 \begin{tabular}{c} Lyapunov \\[-3mm]  exponent  \\[-3mm]  (Gbit/s)   \end{tabular}  & 0.14  &  0.89  &  4.5   &  5.0  &  6.9  \\
\end{tabular}
\end{ruledtabular}
\end{table*}

First, we have extracted a quality factor (Q factor) of  for the interspike time interval from each time trace. 
In Figs. 6(a), (b) and (d) (In Figs. 6(c) and (e)), 
each time interval where $R \leq 150$ $\Omega$ ($R \geq 150$ $\Omega$) is defined as the interspike time interval. 
The Q factor is evaluated as $T_{I}/\sigma_{I}$ during 10 sets of switching of m1 and m2 
where $T_{I}$ ($\sigma_{I}$) is the average value (standard deviation) of interspike time interval.
As we see the enhanced chaotic magnetization dynamics in the in-plane magnetized spin valve in Sec. V, 
the Q factor decreases when the anisotropy changes from out-of-plane 
((a) SV$_{\rm OOP1}$ and (b) SV$_{\rm OOP2}$) to in-plane ((c) SV$_{\rm IP1}$, (d) SV$_{\rm IP2}$ and (e) SV$_{\rm IP3}$).  
In (a) SV$_{\rm OOP1}$, the Q factor exceeds our analyzable upper limit of $10^{4}$ 
because $\sigma_{I} = 0$ in our simulation with a time step of 1 ps. 
The Q factor also decreases by the introduction of dipolar-field interaction 
between layers ([(a) SV$_{\rm OOP1}$ v.s. (b) SV$_{\rm OOP2}$] and 
[(c) SV$_{\rm IP1}$ v.s. (d) SV$_{\rm IP2}$ and (e) SV$_{\rm IP3}$]). 
However, the flux-closure structure in Fig. 6(d) hardly improves the Q factor. 
Nevertheless, it recovers the full amplitude of resistance oscillation compared to Fig. 6(e) and reduces the threshold current $J_{th}$.

Then we have conducted Recurrence Quantification Analysis of $R(t)$ for each structure. 
Recurrence plots for each structure are shown at the bottom of Fig. 6. 
A recurrence plot 
\cite{eckmann_recurrence_1987, marwan_recurrence_2007}
%[J. P. Eckmann, Europhys. Lett. (1987)] [N. Marwan, Phys. Rep. (2007)] 
is a square matrix, 
in which the matrix elements correspond to those times at which a similar resistance state recurs, i.e., 
a plot of 
${\mathfrak R_{\iota,\kappa}}=\Theta(\epsilon_{\iota} - \| R(t_{\iota}) - R(t_{\kappa} ) \| )$. 
Here, $t_{\iota}$ and $t_{\kappa}$ are time during about 10 periods of resistance oscillation shown in the middle panels of Fig. 6. 
$\epsilon_{\iota}$ is a threshold distance, and $\epsilon_{\iota} = 0.05$ $\Omega$ is chosen in Fig. 6.  
$\Theta$ is the Heaviside function, and the elements where ${\mathfrak R_{\iota,\kappa}}=1$ are dots in the recurrence plots. 
In other words, the elements where $R(t_{\iota}) \sim R(t_{\kappa})$ appear as dots in the plots. 
Trivial dots at the matrix diagonal elements at $t_{\iota}= t_{\kappa}$ are removed. 
A perfectly periodic 
oscillator will have dots mainly along the diagonal. 
In Figs. 6(d) and (e), the plots show patterns with reduced regularity reflecting their high degree of chaos compared to the cases of Figs. 6(a)-(c).

Results of Recurrence Quantification Analysis 
\cite{webber_dynamical_1994, marwan_nonlinear_2002, marwan_recurrence_2007},
%[C. L. Webber Jr.,and J. P.Zbilut, J Appl Physiol. (1994)] [N. Marwan and J. Kurths,  Phys. Lett. A, (2002)] [N. Marwan, Phys. Rep. (2007)], 
i.e., $DET$, $L$, $L_{\rm max}$ and $ENTR$ are summarized in the middle of TABLE III. 
$DET$, $L$, $L_{\rm max}$, and $ENTR$ are quantities characterized by the diagonal lines in a recurrence plot. 
The lengths of diagonal lines are directly related to the ratio of 
predictability inherent to the system. 
Suppose that the states at times ι and κ are neighboring. 
If the system 
exhibits predictable behavior,
similar situations will lead to a similar future, 
i.e., the probability for $R(t_{\iota}) \sim R(t_{\kappa})$  is high. 
For perfectly 
periodic
systems, this leads to infinitely long diagonal lines. 
In contrast, if the system is 
chaotic,
the probability for $R(t_{\iota}) \sim R(t_{\kappa})$  will be small 
and we only find single points or short lines. 
In accordance with the evaluated Q factors, 
$DET$, $L$, $L_{\rm max}$, and $ENTR$ decreases when the anisotropy changes from out-of-plane to in-plane.  
They also decrease by the introduction of dipolar-interaction between layers.

Then we have determined the Lyapunov exponent from each time trace \cite{wolf_determining_1985}. 
The Lyapunov exponent is a quantity that characterizes the rate of separation of infinitesimally close trajectories in dynamic systems. 
Lyapunov exponents were evaluated with about 100 periods of resistance oscillation for each structure. 
As we have expected, the Lyapunov exponent, characterizing the degree of chaos, increases 
when the anisotropy changes from out-of-plane to in-plane. 
It also increases in the in-plane configuration 
when the strength of dipolar interaction between layers increases. 
These results show that the degree of chaos can be tuned in a wide range 
by engineering the magnetic stack and anisotropies, which is suitable for various
neuromorphic computing applications.

%========================================
% Tuning chaos by current
%========================================
\section{Tuning chaos by current}
We also checked the tunability of chaos by current.  
The evaluated current density dependence of quality factors (Q factors) and Lyapunov exponents are shown in Fig. 7.  
$J/|J_{th}|$ represents the current density normalized by the threshold current density for each polarity of current. 
The Lyapunov exponents at $J/|J_{th}|= \pm1$ are not shown 
because the too long interspike time interval against pulse width makes evaluation of Lyapunov exponent itself impossible 
and long simulations of 100 periods are not possible with our computational capacity. 
Both trends in Figs. 7(a) and (b) show that the degree of chaos is increased by increasing the magnitude of $J/|J_{th}|$. 
A cause of the increased degree of chaos at large $|J/J_{th}|$ can be the increased instability of m2 during the interspike time interval. 
Fluctuations of m2 strongly vary the angle between magnetizations $\hat{\theta}_{12}$ that gives the torque strength. 
Therefore, the switching of m1 will be complex through the dynamics of m2. 
The trend in Fig. 7 means that the degree of chaos can be tuned in a wide range by the dc current. 
The tunability of chaos by current is quite beneficial because it enables the control of chaos in real-time in a ready-made circuit.

%==============================
% Fig. 7
%==============================
\begin{figure}[H]
%\begin{figure}[P]
\includegraphics [width=0.95\columnwidth] {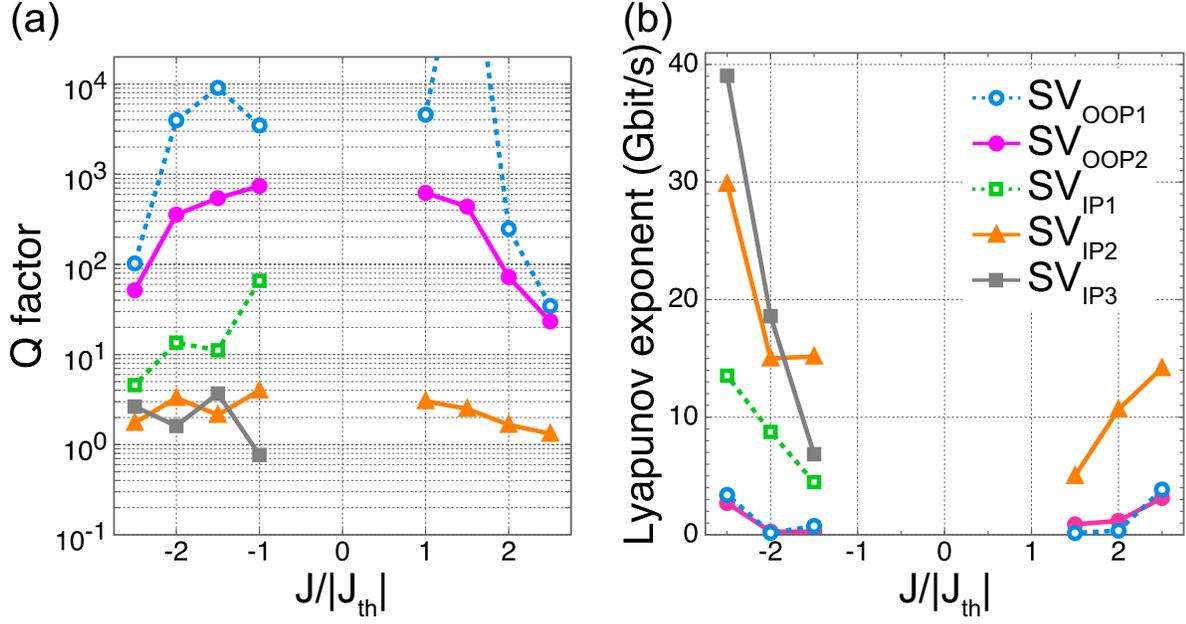}
\caption{\label{fig:fig7} 
(a) $J/|J_{th}|$ dependence of Q factor of interspike time interval and 
(b) Lyapunov exponent. Open circles on dotted lines are for SV$_{\rm OOP1}$. 
Solid circles on solid lines are for SV$_{\rm OOP2}$. Open squares on dotted lines are for SV$_{\rm IP1}$.  
Solid triangles on solid lines are for SV$_{\rm IP2}$. Solid squares on solid lines are for SV$_{\rm IP3}$. 
}  
\end{figure}

%========================================
%  Conclusion
%========================================

\section{conclusion}
We proposed a simple way to imitate neuron spiking in high-magnetoresistance nanoscale spin valves 
where both magnetic layers are free and can be switched by spin torque. 
Our numerical-simulation results show 
that the windmill motion induced by spin torque in the proposed spintronic neurons 
gives rise to spikes whose shape and frequency can be tuned through the amplitude of injected dc current. 
We also found that these devices can exhibit chaotic oscillations. 
By evaluating the quality factors of interspike time intervals and Lyapunov exponents, 
as well as conducting Recurrence Quantification Analysis for the time evolutions of resistance, 
we demonstrate that the degree of chaos can be tuned in a wide range 
by engineering the magnetic stack and anisotropies 
and by changing the dc current. 
The degree of chaos increases 
when the anisotropy of the free layer changes from out-of-plane to in-plane. 
It also increases when the dipolar-field interaction between the free layers increases. 
The proposed spintronic neuron is a promising building block 
for hardware neuromorphic chips leveraging complex non-linear dynamics for computing.

\acknowledgements
This work was partly
supported by JSPS KAKENHI Grant No. JP16K17509 and
the European Research Council (ERC) under grant bioSPINspired 682955.

%========================================
% APPENDIX  MODEL IN MICROMAGNETIC SIMULATIONS
%========================================
\appendix
\section{MODEL IN MICROMAGNETIC SIMULATIONS}
\label{sec:appendix}
In micromagnetic simulations, ${\bm m}_{i} = ( m_{xi}$, $m_{yi}$, $m_{zi})$ of Eqs. (\ref{eq:LLG1}) and (\ref{eq:LLG2})  mean the unit magnetization vector of a unit cell at the position ${\bm r}_{i}$, ${\bm m}_{i}({\bm r}_{i})$.  
The simulations were conducted with the simulation code, 
SpinPM \cite{Spin-PM}. 
%[Ref. Spin-PM]. 
In micromagnetic simulations of this article, 
each magnetic layer is divided into unit cells with the area of 4 nm$ \times $4 nm. 
In the third term on the right side of Eqs. (\ref{eq:LLG1}) and (\ref{eq:LLG2}), i.e., the Slonczewski-torque term, 
$x$ and $y$ components of ${\bm r}_{1}$ and ${\bm r}_{2}$ are the same.

In Eqs. (\ref{eq:LLG1}) and  (\ref{eq:LLG2}), $\textbf{\textit{H}}_{\rm eff}$ is the effective field expressed as
\begin{align}
  \label{eq:HeffMicromag}
\textbf{\textit{H}}_{\rm eff} = \textbf{\textit{H}}_{\rm exch} + \textbf{\textit{H}}_{\rm anis} + \textbf{\textit{H}}_{\rm dip} + \textbf{\textit{H}}_{\rm RKKY}. 
\end{align}
$\textbf{\textit{H}}_{\rm exch}$ represents the exchange field expressed as
\begin{align}
  \label{eq:Hexch}
\textbf{\textit{H}}_{\rm exch}=\frac{2A}{\mu_{0}M_{\rm s}} \nabla^{2} \textbf{\textit{m}}.
\end{align}
$A$ is the exchange constant. 
In the micromagnetic simulations, it is assumed to be $A = 2\times10^{-11}$ J/m in this article. 
$\textbf{\textit{H}}_{\rm dip}$ represents the dipolar field.  
$\textbf{\textit{H}}_{\rm dip}$ on the position ${\bm r}$ is expressed as
\begin{align}
  \label{eq:HdipMicromag}
\textbf{\textit{H}}_{\rm dip} &=  -
 \frac{M_{\rm s}}{4 \pi} \int_{Vol} 
\left[ 
  \frac{\textbf{\textit{m}}(\textbf{\textit{r}}^{\prime})}{|\textbf{\textit{r}} - \textbf{\textit{r}}^{\prime}|^{3} }  \right.  \nonumber\\
   &\hspace{22mm} - \left.
\frac{3 \left[\textbf{\textit{m}}(\textbf{\textit{r}}^{\prime}) \cdot (\textbf{\textit{r}} - \textbf{\textit{r}}^{\prime}) \right] } 
  {|\textbf{\textit{r}} - \textbf{\textit{r}}^{\prime}|^{5}} 
(\textbf{\textit{r}} - \textbf{\textit{r}}^{\prime}) 
  \right]   d\textbf{\textit{r}}^{\prime}.
\end{align}
%\begin{align}
%  \label{eq:HdipMicromag}
%\textbf{\textit{H}}_{\rm dip} =  -
% \frac{M_{\rm s}}{4 \pi} \int_{V} 
%\left[ 
%\frac{\textbf{\textit{m}}(\textbf{\textit{r}}^{\prime})}{|\textbf{\textit{r}} - \textbf{\textit{r}}^{\prime}|^{3} }   
%- 
%\frac{3 \left[\textbf{\textit{m}}(\textbf{\textit{r}}^{\prime}) \cdot (\textbf{\textit{r}} - \textbf{\textit{r}}^{\prime}) \right] } 
%  {|\textbf{\textit{r}} - \textbf{\textit{r}}^{\prime}|^{5}} 
%(\textbf{\textit{r}} - \textbf{\textit{r}}^{\prime})
%\right] d\textbf{\textit{r}}^{\prime}.
%\end{align}
Here the integral is performed over the volume ($Vol$) including all magnetic layers. 
$\textbf{\textit{H}}_{\rm RKKY}$ represents the RKKY coupling field expressed as
\begin{align}
  \label{eq:HRKKY}
\textbf{\textit{H}}_{\rm RKKY}=\frac{J_{\rm RKKY}}{\mu_{0}M_{\rm s}d} \nabla \left(\textbf{\textit{m}} \cdot \textbf{\textit{m}}_{{\rm A}} \right).
\end{align}
Here, $J_{\rm RKKY}$ is the exchange coupling constant. $\textbf{\textit{H}}_{\rm RKKY}$ is considered only in the spin valve shown in Fig. 6(d). ${\bm m}_{\rm A}$ represents the unit magnetization vector of a ferromagnetic layer which is antiferromagnetically-coupled with m, and $J_{\rm RKKY} = -0.1$ mJ/m$^{2}$ is assumed. 
In the scalar product, $x$ and $y$ components of ${\bm r}$ in ${\bm m}({\bm r})$ and ${\bm r}_{\rm A}$ in ${\bm m}_{\rm A}({\bm r}_{\rm A})$ are the same.

%\nocite{*}
%\bibliographystyle{apsrev4-1}
%\bibliography{ref}% Produces the bibliography via BibTeX.

%merlin.mbs apsrev4-1.bst 2010-07-25 4.21a (PWD, AO, DPC) hacked
%Control: key (0)
%Control: author (0) dotless jnrlst
%Control: editor formatted (1) identically to author
%Control: production of article title (0) allowed
%Control: page (1) range
%Control: year (0) verbatim
%Control: production of eprint (0) enabled
%

\end{document}